\documentclass[pra,aps, showpacs, twocolumn, a4paper,tightenlines,balancelastpage]{revtex4}
\usepackage{amsmath,amsfonts,amssymb,mathrsfs,bm}
\usepackage{verbatim,psfrag,times,CJK}
\usepackage{graphicx,graphics,color,epsfig}
\usepackage{flafter}

\usepackage{indentfirst}
\begin{document}

\newcommand{\ket}[1]{| #1 \rangle} 
\newcommand{\bra}[1]{\langle #1 |}

\title{Generating two-photon entangled states in a driven two-atom system} 
\author{Khulud \surname{Almutairi}$^{a,b}$}
\author{Ryszard \surname{Tana\'{s}}$^{c}$}
\email{tanas@kielich.amu.edu.pl}
\author{Zbigniew \surname{Ficek}$^{d}$}
\email{zficek@kacst.edu.sa}
\affiliation{$^{a}$Institute for Quantum Information Science, University of Calgary, Calgary, Alberta T2N 1N4, Canada\\ 
$^{b}$Department of Physics, King Saud University, Riyadh 11451, Saudi Arabia\\
$^{c}$Nonlinear Optics Division, Faculty of Physics, Adam Mickiewicz University, Umultowska 85, 61-614 Pozna\'n, Poland\\
$^{d}$The National Centre for Mathematics and Physics, KACST, P. O. Box 6086, Riyadh 11442, Saudi Arabia}

\date{\today}

\begin{abstract}
We describe a mechanism for a controlled generation of a pure Bell state with correlated atoms that involve two or zero excitations. The mechanism inhibits transitions into singly excited collective states of a two-atom system by shifting them from their unperturbed energies. The shift is accomplished by the dipole-dipole interaction between the atoms. The creation of the Bell state is found to be dependent on the relaxation of the atomic excitation. When the relaxation is not present or can be ignored, the state of the system evolves harmonically between a separable to the maximally entangled state. We follow the temporal evolution of the state and find that the concurrence can be different from zero only in the presence of the dipole-dipole interaction. Furthermore, in the limit of a large dipole-dipole interaction, the concurrence reduces to that predicted for an $X$-state of the system. A general inequality is found which shows that the concurrence of an $X$-state system is a lower bound for the concurrence of the two-atom system. With the relaxation present, the general state of the system is a mixed state that under a strong dipole-dipole interaction reduces the system to an $X$-state form. We find that mixed states admit of lower level of entanglement, and the entanglement may occur over a finite range of time. A simple analytical expression is obtained for the steady-state concurrence which shows that there is a threshold value for the dipole-dipole interaction relative to the Rabi frequency of the driving field above which the atoms can be entangled over the entire time of the evolution.  
\end{abstract}

\pacs{03.65.Yz, 03.67.Bg, 42.50.Dv, 42.50.Hz}

\maketitle

\section{Introduction}

It has been known for many years that two types of maximally entangled Bell states can be generated in a system composed of two two-level atoms; the so-called spin anti-correlated states that are linear superpositions of single excitation product states, and spin correlated states that are linear superpositions of double and zero excitation product states. Recent theoretical and experimental work has demonstrated that single excitation Bell states can be generated in a system of two Rydberg atoms by the dipole-dipole blockade mechanism~\cite{cp10}. The mechanism is often referred to as Rydberg blockade and bears on the elimination of the simultaneous excitation of the atoms by shifting the double-excitation states of the system, an effect attributed to the long range dipole-dipole interaction characteristics of Rydberg atoms~\cite{lf01,uj09,ga10,ge10,hh10,zm10,nm10,ss11,jr10}. Rydberg atoms are highly excited atoms which have large sizes and therefore can have huge dipole moments, proportional to $n^{2}$, where $n$ is the principal quantum number~\cite{ga94}. Because the dipole moments of Rydberg atoms are so large, the atoms can strongly interact with each other even at large distances. 

The physical origin of the dipole-dipole blockade is explained clearly in terms of collective states of the two-atom system. These states provide a more natural basis for interacting atoms~\cite{dic,le70,ag74,ft02}. The effects of the dipole-dipole interaction include in general the creation of maximally entangled single excitation states and the shift of the states from the single-atom energy~\cite{tr82,va92,hs02}.
When the double-excitation states are shifted form their resonances, the two-photon excitation of the system by a resonant laser field becomes suppressed. The two-photon excitation is suppressed without destroying the one-photon excitation. The blockade effect is a simple process for creation of single-excitation entangled states, and the preparation of the entangled states via the dipole blockade has recently been demonstrated experimentally~\cite{gm09,hr07,zi10,wg10}. The presence of only single excitations of the system is manifested by the photon antibunching effect, which signifies that at given time only a single photon is emitted by the system~\cite{cw76,km76,kd77,ft84}. In the experiment of Ga\"etan {\it et al.}~\cite{gm09}, a modification of the Rabi frequency of the driving laser field by $\sqrt{2}$ has been observed that is recognised as a signature of not only the dipole-dipole blockade effect, but also of the creation of a single excitation entangled state. In an earlier dipole blockade experiment, Heidemann~{\it et al.}~\cite{hr07} have observed the characteristic $\sqrt{N}$ scaling of the Rabi oscillations between the ground state and the single excitation multi-atom entangled state of a laser driven mesoscopic ensemble of $N$ ultracold Rydberg atoms. In the experiments of Zhang {\it et al.}~\cite{zi10} and Wilk {\it et al.}~\cite{wg10}, an entanglement between two Rydberg atoms has been demonstrated by measuring the state fidelity of $F = 0.71$ and $F = 0.75$, respectively, that are well above the threshold of $F = 0.5$ required for quantum entanglement. The experiments clearly demonstrate that the dipole-dipole blockade can deterministically generate entangled states in an atomic system. 

The notion of blockade of multi-photon excitations is not restricted to the dipole-dipole blockade, but has been extended to photon blockade for the transport of light through an optical system~\cite{is97,bb05,bi11}, and Coulomb blockade of resonant transport of electrons through small metallic or semiconductor devices~\cite{al86}.
The photon blockade mechanism, similar to the dipole-dipole blockade, prevents absorption of more than one photon by an optical system. In analogy, the Coulomb blockade prevents transport of more than one electron through a metallic or semiconductor device.

The dipole-dipole blockade mechanism described above applies to the process of the creation of maximally entangled single excitation Bell states only. It is the purpose of this paper, therefore, to propose a mechanism which might be useful in achieving entangled two-photon Bell states in a system composed of two coherently driven two-level atoms. As we have already mentioned, these states are linear superposition of product states with double and zero excitations. We shall demonstrate that the states could also be generated with the help of the dipole-dipole interaction between the atoms. Specifically, we analyze the dipole-dipole interaction as a blockade effect for a single photon absorption of the laser field. We work in the collective basis of the system, and find that in the limit of a large dipole-dipole interaction, the collective four level system reduces to an effective two-level two-photon system. The state of this reduced system is then obtained analytically and the nature of the state is fully analyzed. We show that the creation of a pure two-photon  Bell state is dependent on the relaxation of the atomic excitation. When the relaxation is not present or can be ignored, the state of the system evolves harmonically between a separable to the maximally entangled two-photon Bell states. With the relaxation present, the general state of the system is a mixed state that under a strong dipole-dipole interaction reduces the system to an $X$-state form. We find that mixed states admit of lower level of entanglement, and show that there is a threshold value for the dipole-dipole interaction relative to the Rabi frequency of the driving field above which the atoms can be entangled over the entire time of the evolution.

The paper is organized as follows. In Sec.~\ref{sec2}, we give a qualitative explanation of the mechanism for the creation of two-photon Bell states in a driven two-atom system with the help of the dipole-dipole interaction. A detailed calculation of the evolution of the system isolated from a dissipative environment is studied in Sec.~\ref{sec3}. Analytical expressions are obtained for the probability amplitudes of the collective states of the system along with the expression for the concurrence. We show how the shift of the single excitation states may lead to a maximally entangled state involving only the ground and double excited states of the system. In Sec.~\ref{sec4}, we present the conditions for the system to reduce to a two-level two-photon system described by the density matrix in an $X$-state form. Section~\ref{sec5} deals with a practically more realistic model in which the atoms interact with a vacuum reservoir. The interaction results in the dissipation of the atomic excitation and coherence, and we examine the effect of the dissipation on the transient evolution and stationary properties of the concurrence. The remarkably simple analytical expression is obtained for the steady-state concurrence. We summarize our results in the concluding Sec.~\ref{sec6}. Finally, in the Appendix, we present the full set of equations of motion for the density matrix elements and their steady-state solutions.

\section{The model}\label{sec2}

We begin with a qualitative explanation of the concept of the generation of two-photon Bell states in a driven two-atom system with the help of the dipole-dipole interaction. Let us consider a system composed of two identical non-overlapping atoms, separated by a distance $r_{12} =|\vec{r}_{2} -\vec{r}_{1}|$ and coupled to each other through the dipole-dipole interaction. Each atom is modelled as a two-level system (qubit) with the ground state~$\ket{g_i}$ and the excited state  $|e_i\rangle$, $(i=1,2)$ separated by a transition frequency~$\omega_{0}$. In the absence of any external fields, e.g. a driving laser field, the Hamiltonian of the system is of the form
\begin{align}
H = \hbar\omega_{0}\left(S_{1}^{z}+S_{2}^{z}\right) + \hbar\sum_{i\neq j=1}^{2}\Omega_{ij}S_{i}^{+}S_{j}^{-} ,\label{e1}
\end{align}
where $S^{+}_{i}=\ket{e_{i}}\bra{g_{i}}$, $S^{-}_{i}=\ket{g_{i}}\bra{e_{i}}$ and $S_{i}^{z}=(\ket{e_{i}}\bra{e_{i}}-\ket{g_{i}}\bra{g_{i}})/2$ are, respectively, the raising, lowering and energy difference operators of the $i$th atom, and $\Omega_{ij}$ is the dipole-dipole potential between the atoms. The potential depends on the distance between the atoms, mutual orientation of the atomic transition dipole moments, and the orientation of the dipole moments in respect to the interatomic axis. For the case of mutually parallel dipole moments, the potential is given by~\cite{le70,ag74,ft02}
\begin{align}
\Omega_{ij} &= \frac{3}{4}\gamma\left\{ -\left(1-\cos^{2}\theta \right) 
\frac{\cos\!\left( kr_{ij}\right)}{kr_{ij}}\right.  \nonumber \\
&\left. +\left( 1 - 3\cos^{2}\theta\right)\left[\frac{\sin\!\left(kr_{ij}\right) }
{\left(kr_{ij}\right) ^{2}} + \frac{\cos\!\left(kr_{ij}\right) }{\left( kr_{ij}\right) ^{3}}\right]\!\right\} ,\label{e2}
\end{align}
where $\gamma$ is the spontaneous emission rate of the atomic excitations, assumed the same for both atoms, $\theta$ is the angle between the dipole moments of the atoms and the direction of the inter-atomic axis, $k=\omega_{0}/c$, and $r_{ij}$ is the distance between the atoms.

\begin{figure}[h]
\includegraphics[width=0.95\columnwidth]{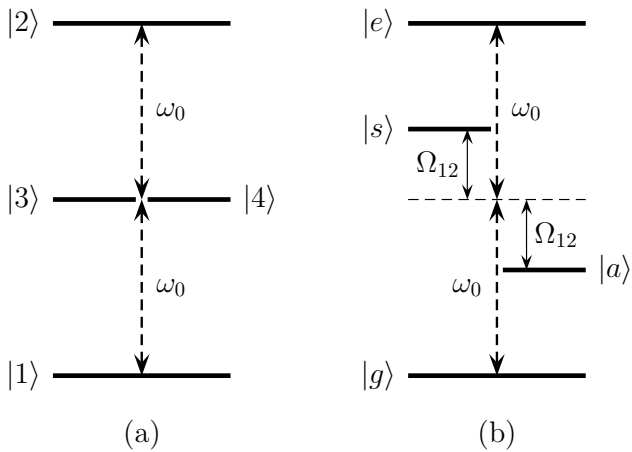}
\caption{Schematic illustration of the idea of the creation of two-photon entangled states in a driven two-atom system. (a) In the absence of the dipole-dipole interaction between the atoms, $\Omega_{12}=0$, the driving laser field of frequency $\omega_{L}=\omega_{0}$ is on resonance with both, one and two-photon transitions of the two-atom system. (b) In the presence of the dipole-dipole interaction, $\Omega_{12}\neq 0$, and then the intermediate states are shifted from the one-photon resonance leaving only the two-photon transition on resonance with the driving field frequency.}
\label{dfig1}
\end{figure}

We can write the Hamiltonian (\ref{e1}) in a matrix form using basis states of two non-interacting atoms that are four product states
\begin{align}
  |1\rangle &= |g_{1}\rangle\otimes|g_{2}\rangle ,\quad |2\rangle = |e_{1}\rangle\otimes|e_{2}\rangle ,\nonumber\\
  |3\rangle &= |g_{1}\rangle\otimes|e_{2}\rangle ,\quad |4\rangle = |e_{1}\rangle\otimes|g_{2}\rangle ,\label{e3}
\end{align}
and find
\begin{align}
  H &= \hbar \left(
    \begin{array}{cccc}
      -\omega_{0}&0&0&0\\
      0&\omega_{0}&0&0\\
      0&0&0&\Omega_{12}\\
      0&0&\Omega_{12}&0
    \end{array}\right) .\label{e4}
\end{align}
Notice that in the absence of the dipole-dipole interaction, the single excitation states $\ket 3$ and $\ket 4$ are degenerate in energy with $E_{3}=E_{4}=0$, and are separated from the zero $\ket 1$ and double $\ket 2$ excitation states by~$\hbar\omega_{0}$, as illustrated in Fig.~\ref{dfig1}(a).

In the presence of the dipole-dipole interaction, the Hamiltonian (\ref{e4}) is not diagonal, and a diagonalization results in the so-called collective or Dicke states~\cite{dic,le70,ag74,ft02}
\begin{align}
  |g\rangle&=|g_{1}\rangle\otimes|g_{2}\rangle ,\quad |e\rangle = |e_{1}\rangle\otimes|e_{2}\rangle ,\nonumber\\
  |s\rangle&=\frac{1}{\sqrt{2}}(|e_{1}\rangle\otimes|g_{2}\rangle+|g_{1}\rangle\otimes|e_{2}\rangle) ,\nonumber\\
  |a\rangle&=\frac{1}{\sqrt{2}}(|e_{1}\rangle\otimes|g_{2}\rangle-|g_{1}\rangle\otimes|e_{2}\rangle) ,\label{e5}
\end{align}
with the corresponding energies
\begin{align}
  E_{g} = -\hbar\omega_{0}  ,\, E_{e}=\hbar\omega_{0} ,\, E_{s}=\hbar\Omega_{12} ,\, E_{a} = -\hbar\Omega_{12} .\label{e6}
\end{align}
We see that the dipole-dipole interaction lifts the degeneracy of the single excitation product states and leads to two  non-degenerate maximally entangled states $\ket s$ and $\ket a$ separated in energy by $2\hbar\Omega_{12}$, as illustrated in Fig.~\ref{dfig1}(b).

We are now in a position to explain qualitatively the idea of the creation of two-photon entangled states with the help of the dipole-dipole interaction. The idea is illustrated in Fig.~\ref{dfig1}, in which the schematic energy-level diagram of a two-atom system is shown. Independent of the basis used, a system composed of two two-level atoms is equivalent to a single four-level system with a ground state, two intermediate states separated from the ground state by energy $\hbar\omega_{0}$ and an upper state separated from the ground state by energy $2\hbar\omega_{0}$. 
Assume that the two-atom system is driven by an external laser field of frequency $\omega_{L}$ resonant with the atomic transition frequency, i.e. $\omega_{L}=\omega_{0}$. Consider separately two cases, the absence and the presence of the dipole-dipole interaction between the atoms. In the absence of the dipole-dipole interaction, the laser field drives on resonance both, the one- and two-photon transitions of the system, as seen from Fig.~\ref{dfig1}(a). In this case, all states of the system are populated by the laser field. Since the one-photon coupling dominates over the two-photon coupling, the intermediate states $\ket 3$ and~$\ket 4$ are more populated than the upper state~$\ket 2$. The situation changes in the presence of the dipole-dipole interaction. The interaction shifts the intermediate states from the one-photon resonance, as shown in Fig.~\ref{dfig1}(b). As a consequence, the one-photon transitions become off resonance to the driving field frequency, but the two-photon transition remains on resonance. Thus, in the presence of the dipole-dipole interaction, a resonant laser field effectively couples only to the two-photon transition of the two-atom system. 

A more quantitative attempt is presented below, where we show that the shift of the intermediate states by the dipole-dipole interaction can lead to the population of the upper state~$\ket e$ without populating the intermediate states $\ket s$ and $\ket a$, which then may result in a pure maximally entangled two-photon state of the system.

\section{Creation of two-photon entangled states}\label{sec3}

Let us first consider the case when the atoms are isolated from the environment. In this case, there is no relaxation of the atomic excitation, but there could still exist a non-zero dipole-dipole coupling between the atoms. In other words, there are no losses due to spontaneous emission and therefore the evolution of the system can be determined by the evolution of a pure state of the system. In practical terms, it could be done by placing the atoms inside separate cavities and to arrange the coupling between the cavities through an optical waveguide~\cite{pf08,lg09}. Alternatively, one could trap single ions or atoms inside separate potential wells and use an additional trapped ion as antennae to enhance the dipole-dipole coupling~\cite{hl10}.

Assume that the two-atom system described by the Hamiltonian (\ref{e1}) is subjected to a continuous driving by an external coherent laser field. The strength of the driving is characterized by the Rabi frequency $\Omega_{0}$. The driving laser field has a traveling-wave character and is propagating in the direction orthogonal to the interatomic axis. The orthogonality of the driving field and the interatomic axis ensues that both atoms experience the same laser field amplitude and phase. The Hamiltonian for this case is given~by
\begin{align}
H &= \hbar\omega_{0}\left(S_{1}^{z}+S_{2}^{z}\right) + \hbar\sum_{i\neq j=1}^{2}\Omega_{ij}S_{i}^{+}S_{j}^{-} \nonumber\\
&-\frac{1}{2}\hbar\Omega_{0}\left[(S_{1}^{+}+S_{2}^{+}){\rm e}^{-i\omega_{L}t} + (S_{1}^{-}+S_{2}^{-}){\rm e}^{i\omega_{L}t}\right] ,\label{eq:1}
\end{align}
where $\omega_{L}$ is the laser field frequency. The laser-atoms coupling part of the Hamiltonian retains only the terms which play a dominant role in the rotating-wave approximation (RWA). Anti-resonant terms which would make much smaller contributions have been omitted. 

When the relaxation of the atomic excitation is not present or can be neglected, the time evolution of the system is governed by the Schr\"odinger equation
\begin{align}
  \label{eq:3}
  i\hbar\frac{d\,|\Psi_{s}(t)\rangle}{dt} = H|\Psi_{s}(t)\rangle ,
\end{align}
where $|\Psi_{s}(t)\rangle$ is the wave function of the driven two-atom system. 

Since the Hamiltonian (\ref{eq:1}) depends explicitly on time, we make the unitary transformation
\begin{align}
  |\Psi(t)\rangle = \exp\left[i\left(H_{0}/\hbar\right)t\right]|\Psi_{s}(t)\rangle ,\label{eq:6}
\end{align}
where
\begin{align}
  \label{eq:5}
  H_{0} = \hbar\omega_{L}(S_{1}^{z}+S_{2}^{z}) ,
\end{align}
and obtain the Schr\"odinger equation for the transformed state
\begin{align}
  \label{eq:7}
  i\hbar\frac{d\,|\Psi(t)\rangle}{dt} = \tilde{H}|\Psi(t)\rangle ,
\end{align}
with the transformed Hamiltonian $\tilde{H}$ of the form
\begin{align}
\tilde{H} &= \hbar\Delta(S_{1}^{z}+S_{2}^{z}) +\hbar\Omega_{12}(S_{1}^{+}S_{2}^{-}+S_{2}^{+}S_{1}^{-}) \nonumber\\
& -\frac{1}{2}\hbar\Omega_{0}\left(S_{1}^{+}+S_{2}^{+}+S_{1}^{-}+S_{2}^{-}\right) .\label{eq:8}
\end{align}
Here, $\Delta=\omega_{0}-\omega_{L}$ is the detuning of the laser frequency from the atomic resonance. 

The transformed Hamiltonian~\eqref{eq:8} does not depend on time, so the Schr\"odinger equation (\ref{eq:7}) has the formal solution
\begin{align}
  \label{eq:9}
  |\Psi(t)\rangle&=\exp\left[-i(\tilde{H}/\hbar)t\right]|\Psi(0)\rangle ,
\end{align}
where $|\Psi(0)\rangle$ is the initial state of the system at $t=0$.

In order to find the time evolution of the wave function for a given initial state of the system, we write the Hamiltonian $\tilde{H}$ in the basis of the collective states (\ref{e5}), and find
\begin{align}
  \label{eq:14}
  \tilde{H} =\hbar\left(
    \begin{array}{cccc}
      -\Delta&0&-\tilde{\Omega}&0\\
      0&\Delta&-\tilde{\Omega}&0\\
      -\tilde{\Omega}&-\tilde{\Omega}&\Omega_{12}&0\\
      0&0&0&-\Omega_{12}
    \end{array}\right) ,
\end{align}
where $\tilde{\Omega} = \Omega_{0}/\sqrt{2}$. 

Since the two-photon coherence, which is of the main interest here, attains maximal values for the laser frequency on resonance with the two-photon transition $\ket g \leftrightarrow \ket e$, we put $\Delta=0$ in Eq.~(\ref{eq:14}) and readily find the following eigenvalues (energies) of the Hamiltonian  
\begin{align}
  E_{1} &= -\hbar\left(\Omega -\frac{1}{2}\Omega_{12}\right) ,\quad 
  E_{2} = \hbar\left(\Omega +\frac{1}{2}\Omega_{12}\right) ,\nonumber\\
  E_{3} &=0 ,\quad E_{4} = -\hbar\Omega_{12} ,\label{eq:15}
\end{align}
with the corresponding eigenvectors (energy states)
\begin{align}
  |\psi_{1}\rangle &= \sqrt{\frac{\alpha_{+}}{2}}\,(|e\rangle+|g\rangle) +\sqrt{\alpha_{-}} |s\rangle ,\nonumber\\
  |\psi_{2}\rangle &= -\sqrt{\frac{\alpha_{-}}{2}}\,(|e\rangle+|g\rangle) +\sqrt{\alpha_{+}} |s\rangle ,\nonumber\\
  |\psi_{3}\rangle&=\frac{1}{\sqrt{2}}(|e\rangle-|g\rangle) ,\quad
  |\psi_{4}\rangle = |a\rangle ,\label{eq:16}
\end{align}
where
\begin{align}
  \alpha_{\pm} =\frac{\Omega_{\pm}}{2\Omega} ,\quad {\rm with}\quad 
  \Omega_{\pm}=\Omega\pm \frac{1}{2}\Omega_{12} ,
 \label{eq:16a}
\end{align}
and 
\begin{align}
  \label{eq:16b}
  \Omega =\sqrt{\Omega^{2}_{0} + \frac{1}{4}\Omega_{12}^{2}}\, .
\end{align}
We see from Eq.~(\ref{eq:15}) that the single excitation antisymmetric state~$|a\rangle$ is the eigenstate of the Hamiltonian with eigenvalue~$-\hbar\Omega_{12}$, the double excitation antisymmetric state $\ket{\psi_{3}}$ is the eigenstate with eigenvalue zero, and the remaining states are superpositions of the single and double excitation states with non-degenerate eigenvalues.

\subsection{Time evolution of the concurrence}

We now turn to the problem of determining the form of an entangled state that could be created by the dipole-dipole interaction shift of the single excitation states and the degree of the resulting entanglement. As a measure of entanglement, we choose the concurrence that for a pure state $|\Psi_{s}(t)\rangle$ of a two-atom system is given by
\begin{align}
  {\cal C}(t) = |\langle\Psi_{s}(t)|\tilde{\Psi}_{s}(t)\rangle| ,\label{eq:27}
\end{align}
where
\begin{align}
  \label{eq:28}
  |\tilde{\Psi}_{s}(t)\rangle = \sigma_{y}^{(1)}\otimes\sigma_{y}^{(2)}|\Psi_{s}^{*}(t)\rangle ,
\end{align}
and $\sigma_{y}^{(i)}\, (i=1,2)$ is the Pauli operator for the $i$th atom. 

For the system considered here, the state $|\Psi_{s}(t)\rangle$ is of the form
\begin{align}
  \label{eq:26}
  |\Psi_{s}(t)\rangle = \exp\left[-i(H_{0}/\hbar)t\right]|\Psi(t)\rangle ,
\end{align}
where $|\Psi(t)\rangle$ is given in Eq.~(\ref{eq:9}).

Given the state of the system, it is straightforward to calculate the concurrence. We have two alternative forms for the state $|\Psi_{s}(t)\rangle$, the diagonal states~(\ref{eq:16}) or the collective states~(\ref{e5}). Of these alternatives, we will prefer the collective states as it is very often found that they form very convenient basis states to discuss properties of the concurrence. Therefore, we invert the transformation~(\ref{eq:16}) to find
\begin{align}
  |g\rangle &= \frac{1}{\sqrt{2}}\left(\sqrt{\alpha_{+}}\,|\psi_{1}\rangle -\sqrt{\alpha_{-}}\, |\psi_{2}\rangle-|\psi_{3}\rangle\right) ,\nonumber\\
  |e\rangle &= \frac{1}{\sqrt{2}}\left(\sqrt{\alpha_{+}}\,|\psi_{1}\rangle -\sqrt{\alpha_{-}}\, |\psi_{2}\rangle+|\psi_{3}\rangle\right) ,\nonumber\\
  |s\rangle &= \sqrt{\alpha_{-}}\, |\psi_{1}\rangle + \sqrt{\alpha_{+}}\, |\psi_{2}\rangle ,\quad
  |a\rangle =|\psi_{4}\rangle ,\label{eq:21}
\end{align}
and write the state vector of the system as
\begin{align}
  \label{eq:24}
|\Psi(t)\rangle = C_{g}(t)\ket g + C_{e}(t)|e\rangle + C_{s}(t)|s\rangle + C_{a}(t)|a\rangle ,
\end{align}
where $C_{n}(t)$ is the probability amplitude of the $n$th state. 

The time evolution of the state vector of the system and so the concurrence depend, of course, on the initial state of the system. Consider as the initial state the ground state $|g(0)\rangle = \ket{g_{0}}$, corresponding to no excitation of the system at $t=0$. In terms of the eigenstates~(\ref{eq:16}), the initial state is of the form
\begin{align}
  \label{eq:22}
  \ket{g(0)} =\frac{1}{\sqrt{2}}\left(\sqrt{\alpha_{+}}\, |\psi_{1}(0)\rangle -
    \sqrt{\alpha_{-}}\,|\psi_{2}(0)\rangle-|\psi_{3}(0)\rangle\right) .
\end{align}
From Eqs.~(\ref{eq:9}) and (\ref{eq:15}), we then find that the time evolution of the coefficients $C_{n}(t)$ is of the form
\begin{align}
  C_{g}(t) &= \frac{1}{2}\left(\alpha_{+}{\rm
      e}^{i\Omega_{-}t}+\alpha_{-}{\rm e}^{-i\Omega_{+}t}+1\right) ,\nonumber\\
  C_{e}(t) &=  \frac{1}{2}\left(\alpha_{+}{\rm
      e}^{i\Omega_{-}t}+\alpha_{-}{\rm e}^{-i\Omega_{+}t}-1\right) ,\nonumber\\
  C_{s}(t) &= \frac{1}{2\sqrt{2}}\frac{\Omega_{0}}{\Omega}\left({\rm
      e}^{i\Omega_{-}t}-{\rm e}^{-i\Omega_{+}t}\right)\,. \nonumber\\
  C_{a}(t) &= 0 .\label{eq:25}
\end{align}
Notice that the amplitude of the antisymmetric state is equal to zero for all times. This is due to the fact that for the configuration considered here, the laser field propagating in the direction perpendicular to the inter-atomic axis, the laser couples exclusively to the symmetric state leaving the antisymmetric state completely decoupled from the  driven states. For the initial state $\ket{g_{0}}$, the antisymmetric state is not populated at $t=0$, and therefore the state will remain unpopulated for all $t>0$.

\begin{figure}[h]
\includegraphics[width=0.95\columnwidth]{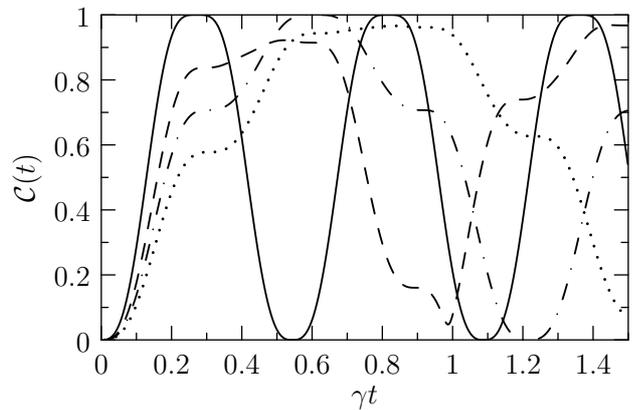}
\caption{Time evolution of the concurrence ${\cal C}(t)$ for $\Omega_{0} =10\gamma$ and different values of the ratio $u = \Omega_{0}/\Omega_{12}$: $u=\sqrt{3/4}$ (solid line), $u= 1.5$ (dashed line), $u=\sqrt{15/4}$ (dashed-dotted line), $u=2.5$ (dotted line).}
\label{dfig2}
\end{figure}

Using Eqs.~(\ref{eq:27}) and (\ref{eq:28}), we find that the solution (\ref{eq:25}) yields the following expression for the concurrence 
\begin{align}
  \label{eq:29}
  {\cal C}(t) &= \left|2C_{g}(t)C_{e}(t)-C_{s}^{2}(t)\right| \nonumber\\
  &=\frac{1}{2}\left|-1+\left(\frac{\Omega_{0}}{\Omega}\right)^{2}\,\text{e}^{-i\Omega_{12}t}\right.\nonumber\\
&\left.\vphantom{\left(\frac{\Omega_{0}}{\Omega}\right)^{2}}+\frac{\Omega_{12}}{2\Omega}{\rm
    e}^{-i\Omega_{12}t}\left(\alpha_{+}\text{e}^{2i\Omega t} -
    \alpha_{-}\text{e}^{-2i\Omega t}\right)\right| .
\end{align}
It is easily verified that the concurrence vanishes when $\Omega_{12}=0$. Thus, crucial for entanglement is the presence of the dipole-dipole interaction that shifts of the single excitation states from their resonant values. The concurrence is different from zero and contains exponentials oscillating  in time with three different frequencies, $\Omega_{12}$, $2\Omega_{-}$, and $2\Omega_{+}$. Since, in general, the three frequencies are not commensurate, the evolution of concurrence is not periodic. However, for some values of the ratio $\Omega_{0}/\Omega_{12}$, the evolution can be periodic. This is shown in Fig.~\ref{dfig2}, where we plot the concurrence versus time $t$ for
several different values of the ratio~$\Omega_{0}/\Omega_{12}$. For $t=0$ the system is separable, ${\cal C}(0)=0$, due to our choice of the initial states of the atoms. Immediately afterwards, the concurrence begins to increase and depending on the ratio $\Omega_{0}/\Omega_{12}$, it oscillates periodically or non-periodically in time. For these values of the ratio $\Omega_{0}/\Omega_{12}$ at which the oscillations are periodical, the concurrence reaches the maximal value of ${\cal C}(t)=1$ that is seen to occur periodically over all times. For the case of non-periodic
oscillations, the maxima are smaller than one. Thus, the effect of the non-periodicity is clearly do decrease the amount of entanglement.

Let us examine the concurrence a little more closely. Consider separately the cases of periodic and non-periodic oscillations. As we have already noticed from Fig.~\ref{dfig2}, and it can also be seen from Eq.~(\ref{eq:29}), for the case of periodic oscillations, the concurrence becomes unity at certain times satisfying the condition
\begin{align}
  \label{eq:30}
  \Omega_{12}t=\pi \quad \text{and}\quad  \Omega t = n\pi ,\quad  n=1,2,\ldots
\end{align}
which can happen only for the discrete values of the ratio
\begin{align}
  \label{eq:32}
  \frac{\Omega_{0}}{\Omega_{12}} = \sqrt{n^2-\frac{1}{4}} ,\quad n=1,2,\ldots
\end{align}
It is easily verified that at the times satisfying the condition (\ref{eq:30}), the probability amplitudes (\ref{eq:25}) reduce to
\begin{align}
C_{g}(t) &= \frac{1}{\sqrt{2}}{\rm e}^{(-1)^{n-1}i\pi/4} ,\quad 
C_{e}(t) = -\frac{1}{\sqrt{2}}{\rm e}^{(-1)^{n}i\pi/4} ,\nonumber\\
C_{s}(t) &= 0 ,\quad n=1,2,\ldots .\label{eq:25a}
\end{align} 
This shows that at these particular times, the system is in a superposition of the ground $\ket g$ and the upper $\ket e$ states with no excitation of the symmetric state $\ket s$.

When Eq.~(\ref{eq:25a}) is used in Eq.~(\ref{eq:24}), we readily find for the state vector of the system of the resulting maximally entangled two-photon Bell state 
\begin{align}
  \label{eq:33}
  |\Psi(t_{n})\rangle = -\frac{\text{e}^{(-1)^{n}i\pi/4}}{\sqrt{2}}\left(|e\rangle +(-1)^{n}i|g\rangle\right) ,\quad n=1,2,\ldots 
\end{align}

We now turn to a detailed analysis of the concurrence for the case of non-periodic oscillations. In this case, the competing effects of one- and two-photon transitions modify the Rabi oscillations and cause increasing distortions of the concurrence. However, we can reduce the destructive effects of the competing transitions by taking large values of the dipole-dipole interaction, $\Omega_{12}\gg \Omega$. It is easily verified from Eq.~(\ref{eq:29}), that in the limit of $\Omega_{12} \gg \Omega$, the amplitudes of the terms oscillating with frequencies $\Omega_{12}$ and $2\Omega_{+}$ are very small, and then the concurrence can be approximated by
\begin{align}
  \label{eq:36}
  {\cal C}(t) \approx \frac{1}{2}\left|-1+\exp\left(-i\frac{2\Omega_{0}^{2}}{\Omega_{12}}t\right)\right| .
\end{align}
We see that at times
\begin{align}
  \label{eq:37}
  t = n\frac{\pi\Omega_{12}}{2\Omega_{0}^{2}} ,\quad n=1,3,5,\ldots
\end{align}
the concurrence is close to its maximal value of ${\cal C}(t)=1$. 

At these times, the state of the system reduces to the maximally entangled two-photon Bell state 
\begin{align}
  \label{eq:38}
  |\Psi(t)\rangle = -\frac{\text{e}^{-i\pi/4}}{\sqrt{2}}\left(|e\rangle -i|g\rangle\right) .
\end{align}
It is seen that the system evolves between the ground $\ket g$ and the doubly-excited state $\ket e$ and at some discrete time can be found in the maximally entangled two-photon Bell state. The maximal entanglement is due to the two-photon coherence. 

It is straightforward to show using Eq.~\eqref{eq:25} that the one- and two-photon coherences evolve in time as 
\begin{align}
  \tilde{\rho}_{gs}(t) &= C_{g}(t)C_{s}^{*}(t)=\frac{1}{4}\frac{\tilde{\Omega}}{\Omega}\left[{\rm e}^{-i\Omega_{-}t}-{\rm e}^{i\Omega_{+}t}\right.\nonumber\\
 &\left.+\alpha_{-}\left({\rm e}^{-2i\Omega t}-1\right)-\alpha_{+}
\left({\rm e}^{2i\Omega t}-1\right)\right]\nonumber\\
  \tilde{\rho}_{es}(t) &= C_{e}(t)C_{s}^{*}(t)=\frac{1}{4}\frac{\tilde{\Omega}}{\Omega}\left[-{\rm e}^{-i\Omega_{-}t}+{\rm e}^{i\Omega_{+}t}\right.\nonumber\\
 &\left.+\alpha_{-}\left({\rm e}^{-2i\Omega t}-1\right)-\alpha_{+}
\left({\rm e}^{2i\Omega t}-1\right)\right]\nonumber\\
  \tilde{\rho}_{ge}(t) &= C_{g}(t)C_{e}^{*}(t)
  = -\frac{1}{2}\left(\frac{\tilde{\Omega}}{\Omega}\right)^{2}\sin^{2}\Omega t  \nonumber\\
  & +\frac{1}{2}i\left(\alpha_{-}\sin\Omega_{+}t-\alpha_{+}\sin\Omega_{-}t\right) ,\label{eq:39}
\end{align}
where $\tilde{\rho}_{ge}(t)= \rho_{ge}(t)\exp(-2i\omega_{0}t)$ is the slowly varying part of the coherence.

It follows from Eqs.~(\ref{eq:32}), (\ref{eq:37}) and (\ref{eq:39}) that the one-photon coherences vanish, $\tilde{\rho}_{gs}(t) =\tilde{\rho}_{se}(t) = 0$, and the two-photon coherence $|\tilde{\rho}_{ge}(t)|$ becomes maximal, $|\tilde{\rho}_{ge}(t)| = 1/2$, at times the concurrence is maximal. Thus, we must conclude that the entanglement is created by the two-photon coherence with no population of the single excitation state $\ket s$.

\subsection{$X$-state as a lower bound for entanglement}\label{sec4}

The results for entanglement created by the two-photon coherence bear interesting relation to an $X$-state entanglement~\cite{ye07}. Therefore, it is interesting to contrast the concurrence considered under the two-photon coherence with the concurrence of the $X$-state system. Since the dipole-dipole interaction shifts the single excitation states from resonance with the driving field, the one-photon coherences are all equal to zero and then the density matrix of the system takes the~$X$-state form
\begin{align}
  \rho(t) = \left(
      \begin{array}{cccc}
        \rho_{gg}(t)&0&0&\rho_{ge}(t)\\
        0&\rho_{aa}(t)&0&0\\
        0&0&\rho_{ss}(t)&0\\
        \rho_{eg}(t)&0&0&\rho_{ee}(t)
      \end{array}\right) . \label{eq:39a}
\end{align}
The concurrence of the system determined by the $X$-state density matrix is readily found to be
\begin{align}
  \label{eq:41}
  {\cal C}_{x}(t) &= 2|\rho_{ge}(t)| -\rho_{ss}(t) \nonumber\\
    &= 2|C_{g}(t)C_{e}^{*}(t)|-|C_{s}(t)|^{2} ,
\end{align}
where we have taken into account that $\rho_{aa}(t)=0$.

We wish to compare the concurrence ${\cal C}_{x}(t)$ with the concurrence ${\cal C}(t)$, Eq.~\eqref{eq:29}, and see whether the concurrence of the system discussed above is larger or smaller than that predicted by the $X$-state system. From Eq.~(\ref{eq:29}), we have
\begin{align}
  \label{eq:42}
  {\cal C}(t) = 2|C_{g}(t)C_{e}(t)-C_{s}^{2}(t)| ,
\end{align}
and then applying the inequality
\begin{align}
  |z_{1}z_{2}-z_{3}^{2}|\ge |z_{1}z_{2}^{*}|-|z_{3}|^{2} ,\label{eq:43}
\end{align}
which holds for arbitrary complex numbers $z_{1},z_{2}$ and $z_{3}$, we readily find that ${\cal C}(t)\ge {\cal C}_{x}(t)$. This inequality always holds true, so we may conclude that values of ${\cal C}_{x}(t)$, which determines concurrence for the $X$-state system, are lower bounds for the concurrence of the present system. In addition, if the phases of the complex amplitudes are chosen such that $\phi_{g}+\phi_{e} - 2\phi_{s} = 0$, the inequality becomes equality. Moreover, for the case of $C_{s}(t) = 0$, that occurs at times the concurrence ${\cal C}(t)$ reaches the maximum ${\cal C}(t)=1$ value, we have ${\cal C}(t) = {\cal C}_{x}(t)$. This means that under the ideal two-photon excitation, the system behaves as an $X$-state system. 
\begin{figure}[h]
\centering
\includegraphics[width=0.95\columnwidth]{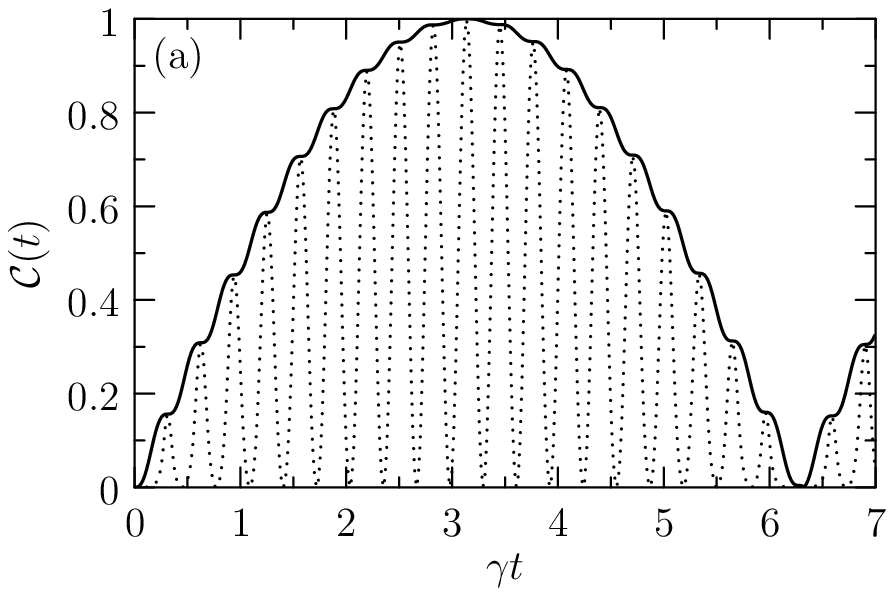}\\
\includegraphics[width=0.95\columnwidth]{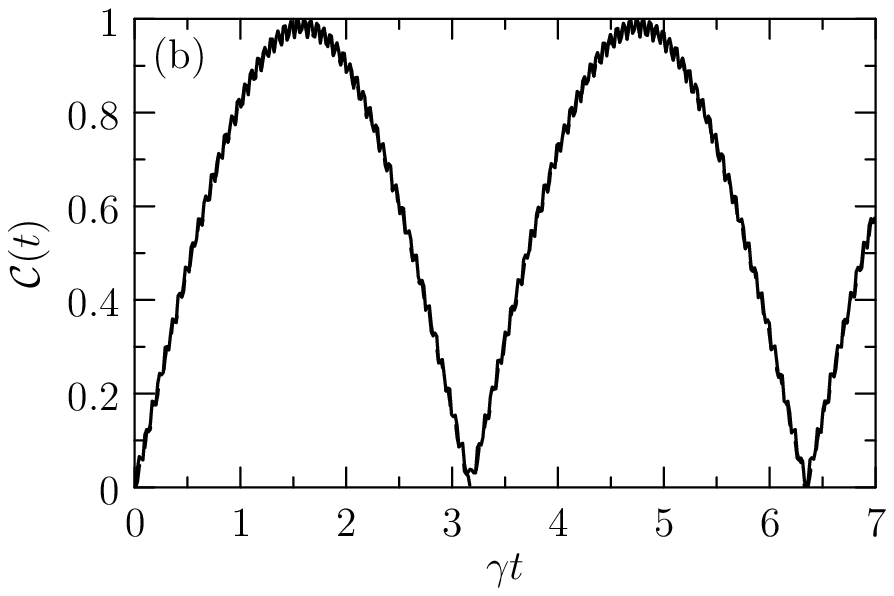}
\caption{Time development of the concurrences ${\cal C}(t)$ (solid line) and ${\cal C}_{x}(t)$ (dashed line) for $\Omega_{0} =10\gamma$ and different values of $\Omega_{12}$: (a) $\Omega_{12} =\gamma$, and (b) $\Omega_{12} =100\gamma$. In the case (b), both concurrences overlap.}
\label{fig:3}
\end{figure}

The above considerations are illustrated in Fig.~\ref{fig:3}, which shows the concurrences ${\cal C}(t)$ and ${\cal C}_{x}(t)$ as a function of time for two different values of $\Omega_{12}$. For a small $\Omega_{12}$, the time evolution of the concurrences exhibits a modulation of different amplitudes, and the effect of an increasing $\Omega_{12}$ is evident in the decrease of the difference between~${\cal C}(t)$ and ${\cal C}_{x}(t)$.

\section{Master equation}\label{sec5}

The above discussion of the creation of the two-photon Bell state describes a simplified case, which ignores dissipative effects of spontaneous emission resulting from the coupling of the atoms to an external multi-mode reservoir. We now extend the model to include the dissipation in the system. In this case, a general state of the system is a mixed state and the dynamics of the atoms are then determined by the evolution of the density operator of the system. The density operator satisfies the master equation which in the interaction picture can be written~as
\begin{align}
\dot{\rho} =& -i\Delta\left[S_{1}^{z}+S_{2}^{z},\rho\right] -i\sum_{i\neq j=1}^{2}\Omega_{ij}\left[S_{i}^{+}S_{j}^{-},\rho\right]\nonumber\\
&+\frac{1}{2}i\Omega_{0}\sum_{i=1}^{2}\left[S_{i}^{+}+S_{i}^{-},\rho\right] \nonumber\\
&-\sum_{i,j=1}^{2}\gamma_{ij}(\rho S_{i}^{+}S_{j}^{-}+S_{i}^{+}S_{j}^{-}\rho-2S_{j}^{-}\rho S_{i}^{+}) .\label{eq:44}
\end{align}
The term of the master equation, involving $\gamma_{ij}$, represents the evolution of $\rho$ due to dissipation in the atomic system. The term contains contributions, proportional to $\gamma_{ii}\equiv \gamma$ representing the damping of the $i$th atom by spontaneous emission, assumed to be independent of~$i$, so that both atoms are equally damped by the field. Apart from the contribution of the individual atoms, the term contains contributions, proportional to $\gamma_{ij}\, (i\neq j)$, that represent the collective damping resulting from the mutual exchange of spontaneously emitted photons through the common reservoir~\cite{dic,le70,ag74,ft02}. The parameter $\gamma_{ij}\, (i\neq j)$ depends on the distance between the atoms and the orientation of the atomic dipole moments in respect to the interatomic axis
\begin{align}
\gamma_{ij} &= \frac{3}{2}\gamma\left\{ \left(1 -\cos^{2}\theta \right)
 \frac{\sin\!\left( kr_{ij}\right)}{kr_{ij}}\right.  \nonumber \\
&\left. + \left( 1- 3\cos^{2}\theta \right)\left[ \frac{\cos\!\left(kr_{ij}\right) }
{\left( kr_{ij}\right) ^{2}}\!-\!\frac{\sin\!\left(kr_{ij}\right) }{\left( kr_{ij}\right) ^{3}}\right]\! \right\} .\label{eq:45}
\end{align}
In general, $\gamma_{ij}$ involves terms oscillating at frequency $kr_{ij}$, multiplied by inverse powers of $kr_{ij}$ ranging from $(kr_{ij})^{-1}$ to $(kr_{ij})^{-3}$. 
Note that $\gamma_{ij}$ is symmetric under the exchange of the atoms, i.e., $\gamma_{21}=\gamma_{12}$. For small atomic separations, $kr_{ij}\ll 1$, the collective damping $\gamma_{ij}$ approaches $\gamma$, whereas for large separations $\gamma_{ij}$ vanishes. The later corresponds to the case where the atoms are independently damped by the reservoir. This case is equivalent to the situation where the atoms are damped by their own independent reservoirs. 

We now employ the master equation (\ref{eq:44}) to find equations of motion for the matrix elements of the atomic density operator $\rho$. We use the collective states basis~(\ref{e5}) as the representation of the density operator. As in the previous sections, we assume that both atoms experience the same amplitude and phase of the driving field and we focus on the two-photon resonance, i.e., we put $\Delta =0$. In so doing, we find that the equations of motion split into two independent sets, one composed of nine inhomogeneous and the other composed of six homogeneous coupled equations. The sets together with their steady-state solutions are listed in Appendix~A.

\subsection{Effect of spontaneous emission on entanglement}

We wish to examine the time development of the concurrence in the presence of spontaneous decay of the atomic excitation and coherence. In this case a general state of the system is a mixed state described by the density operator $\rho$. 

The concurrence of a mixed state of a two-atom system is defined as~\cite{woo}
\begin{eqnarray}
{\cal C} = \max\left(0,\sqrt{\lambda_{1}}-\sqrt{\lambda_{2}}-\sqrt{\lambda_{3}} -\sqrt{\lambda_{4}}\,\right) ,\label{eq:46}
\end{eqnarray}
where $\lambda_{i}$ are the the eigenvalues, putted in decreasing order, of the matrix
\begin{eqnarray}
R=\rho\tilde{\rho} ,\label{eq:47}
\end{eqnarray}
with
\begin{eqnarray}
\tilde{\rho} = \sigma_{y}\otimes\sigma_{y}\rho^{\ast}\sigma_{y}\otimes\sigma_{y} ,\label{eq:48}
\end{eqnarray}
and $\rho^{\ast}$ denotes the complex conjugate of $\rho$.

The concurrence is specified by the density matrix of a given system and thus can be determined from the knowledge of the density matrix elements. Assume that initially, prior the laser field was turned on at $t=0$, the atoms were in their ground states, i.e., $\rho_{gg}(0)=1$ and the other matrix elements equal to zero. 
It is easily verified from Eq.~(\ref{eq:A5}) that in this case, the coherences between the triplet states and the antisymmetric state are equal to zero for all times. As a consequence, the density matrix of the system, written in the collective basis $(\ket g, \ket e, \ket s, \ket a)$, takes the simplified form 
\begin{align}
  \rho(t)&=\left(
    \begin{array}{cccc}
     \rho_{gg}(t)&\rho_{ge}(t)&\rho_{gs}(t)&0\\
      \rho_{eg}(t)&\rho_{ee}(t)&\rho_{es}(t)&0\\
     \rho_{sg}(t)&\rho_{se}(t)&\rho_{ss}(t)&0\\
      0&0&0&\rho_{aa}(t)
    \end{array}\right) .\label{eq:49}
\end{align}

The time evolution of the density matrix elements is found by solving the set of nine coupled differential equations~(\ref{eq:A4}). These equations are cumbersome for an analytical solution because of the coupling between the populations and coherences. Therefore, we use a numerical method to find the evolution of the density matrix elements from which we then compute the time evolution of the concurrence. 

Note that the density matrix (\ref{eq:49}) is not diagonal in the basis of the collective states. This means that, in general, the collective states are not the eigenstates of the system. In principle, it is possible to find the diagonal states simply by direct diagonalization of the matrix~(\ref{eq:49}). However, diagonal states such obtained are complicated in form and thus difficult to interpret. An alternative way is to compare the general state (\ref{eq:49}) with approximate states in which the system could be found under the pure two-photon excitation. 

Of these alternatives, we consider a state described by an approximate density operator of the form 
\begin{align}
\rho(t) = (1-4\rho_{aa}(t))|\Psi(t)\rangle\langle\Psi(t)|+\rho_{aa}(t)\mathbb{I} ,\label{eq:50}
\end{align}
where $\ket{\Psi(t)}$ is the pure state of the system, calculated in Sec.~\ref{sec3}, and $\mathbb{I}$ is the $4\times 4$ identity matrix. 
The approximate density operator represents a state of the system that is an incoherent superposition of the pure state $\ket{\Psi(t)}$ and the antisymmetric state $\ket a$. This choice is suggested by an observation from Eq.~(\ref{eq:A4}) that the spontaneous decay couples the antisymmetric state to the triplet states. This means that this coupling may play an important role in the creation of the two-photon Bell states.

We also compare the general state (\ref{eq:49}) with a state described by the density matrix of the $X$-state form. Therefore, we consider the criterion ${\cal C}_{x}$ for entanglement of an~$X$-state  system given by~\cite{tf04}
\begin{align}
  {\cal C}_{x}(t)&={\rm max}\left\{0,2|\rho_{ge}(t)|\right.\nonumber\\
  &\left. -\sqrt{[\rho_{ss}(t)+\rho_{aa}(t)]^{2}-[2{\rm Re}\rho_{sa}(t)]^{2}}\right\} .\label{eq:55a}
\end{align}
Since for the general state $\rho_{sa}(t)=0$, the criterion (\ref{eq:55a}) simplifies~to
\begin{eqnarray}
  {\cal C}_{x}(t) ={\rm max}\left\{0,2|\rho_{ge}(t)|-\left[\rho_{ss}(t)+\rho_{aa}(t)\right]\right\} .
\end{eqnarray}

We now present some numerical calculations that illustrate the effect of spontaneous emission on the pure two-photon Bell state created with the help of the dipole-dipole interaction. The time development of the concurrence is calculated for the actual state described by the density matrix~\eqref{eq:49} and, for comparison, for an approximate state described by the density matrix~\eqref{eq:50} and also for an $X$ state of the system. 
\begin{figure}[h]
\includegraphics[width=0.95\columnwidth]{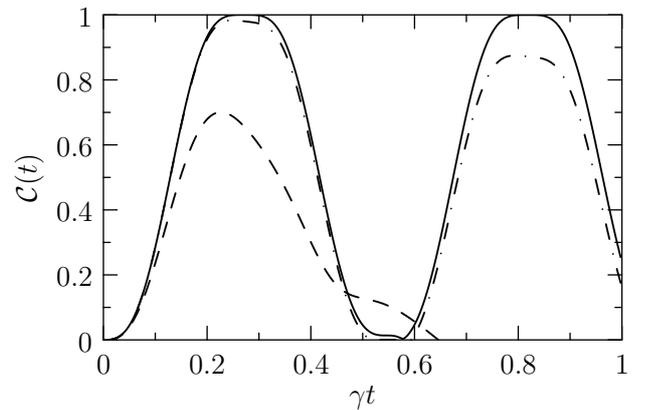}
\caption{Time development of the concurrence for three different states of the system. The solid line shows the concurrence of the pure state of the system decoupled from the reservoir. The dashed line is the concurrence of the actual state of the atoms coupled to a common reservoir, and the dashed-doted line is the concurrence of the approximate state of the system, Eq.~\eqref{eq:50}. In each case $\Omega_{0} =10\gamma$ and $\Omega_{12}=\Omega_{0}/\sqrt{3/4}$. In last two cases, the distance between the atoms is $r_{12}=0.078\lambda$ and the dipole moments are polarized in the direction perpendicular to the interatomic axis, $\theta =\pi/2$. This choice of the parameters has been made to get $\Omega_{12}=11.48\gamma = \Omega_{0}/\sqrt{3/4}$.}
  \label{fig:4}
\end{figure}

Figure~\ref{fig:4} shows the time development of the concurrence of the system of two atoms located at a small distance, $r_{12}=0.078\lambda$, and coupled to a common reservoir. Also shown are concurrences of the system decoupled from the reservoir and that determined by the density matrix~\eqref{eq:50}. It is easy seen that the concurrences evolve in decidedly different ways. In particular, the concurrence of the state approximated by the density matrix~(\ref{eq:50}) does not reproduce the concurrence of the actual state of the system. It is rather close to the concurrence of the pure state of the system of atoms decoupled from the reservoir. The concurrence oscillates in time with the same frequency as that of the isolated system and persists over many Rabi periods. The amplitude of the oscillations decreases slowly in time and is damped out in a time of order $(\gamma -\gamma_{12})^{-1}$. 

The reason for this feature of the approximate state of the system can be understood by referring to the asymmetry introduced to the system by the competing effects of spontaneous emission from the upper state $\ket e$ to the symmetric and antisymmetric states. It is easily verified from Eq.~\eqref{eq:A4} that the antisymmetric state $\ket a$ is populated by spontaneous emission from the state $\ket e$ at a sub-radiant rate $\gamma-\gamma_{12}$, which for small distances between the atoms is much smaller that the superradiant rate, $\gamma +\gamma_{12}$, at which the symmetric state $\ket s$ is populated. This means that the redistribution of the noise in the system is not isotropic. Consequently, the symmetric and antisymmetric states are affected differently by spontaneous emission. A considerable part of the vacuum noise is transferred to the symmetric state with only a small fraction being transferred to the antisymmetric state. Thus, the population~$\rho_{aa}(t)$ is not a good measure of the noise distribution between the states of the system of two atoms coupled to a common reservoir. 
\begin{figure}[h]
   \includegraphics[width=0.95\columnwidth]{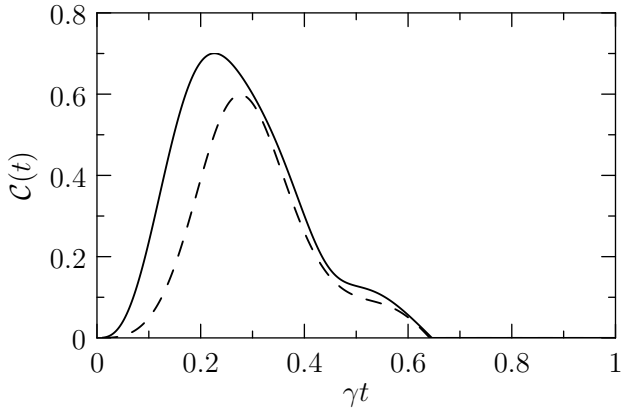}
  \caption{The time evolution of the concurrence of the actual state of the system ${\cal C}(t)$ (solid line) and the approximate $X$ state ${\cal C}_{x}(t)$ (dashed line) for the same parameters as in Fig.~\ref{fig:4}.}
  \label{fig:5}
\end{figure} 

The actual mixed state of the system is much better approximated by the $X$ state density matrix. It is convincingly seen from Fig.~\ref{fig:5}, where we graph the concurrence ${\cal C}(t)$ of the actual state and the concurrence ${\cal C}_{x}(t)$ of the $X$ state of the system for the same parameters as in Fig.~\ref{fig:4}. The concurrences behave in a similar fashion that ${\cal C}_{x}(t)$ remains quite close to~${\cal C}(t)$ for all times. The entanglement lives over a restricted time range with no oscillations present even though the Rabi frequency is high. This is an example of the phenomenon of sudden death of entanglement~\cite{ye04,ey07,zh01}. Notice that~${\cal C}_{x}(t)$ does not exceed the actual concurrence ${\cal C}(t)$, same feature as predicted above for the pure state, see Sec.~\ref{sec4}.
\begin{figure}[h]
  \centering
  \includegraphics[width=0.95\columnwidth]{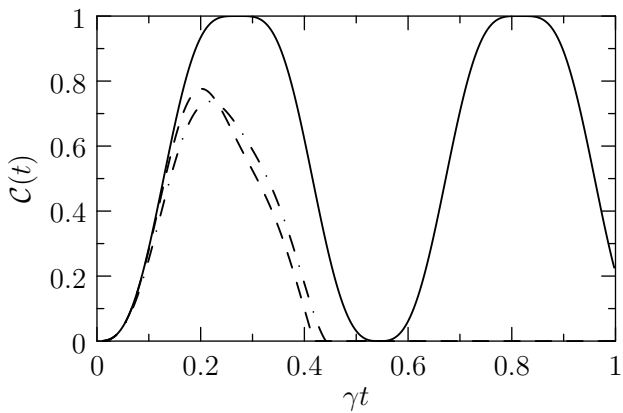}
  \caption{The time evolution of the concurrence of the dipole-dipole interacting atoms for two different arrangements of the coupling to an external reservoir. The solid line is the concurrence of the system decoupled from the reservoir. The dashed line is the concurrence of the actual state of the atoms coupled to independent reservoirs $(\gamma_{12}=0)$, and the dashed-dotted line is the concurrence of the approximate state of the atoms, Eq.~\eqref{eq:50}, and also coupled to independent reservoirs. In each case, $\Omega_{0} =10\gamma$ and $\Omega_{12}=\Omega_{0}/\sqrt{3/4}$.}
  \label{fig:6}
\end{figure}

Although the density matrix \eqref{eq:50} differs considerably from the density matrix of the actual state of the system when the atoms interact with a common reservoir, it does approximate quite well the actual state in the case of independent reservoirs, i.e., when $\gamma_{12}=0$. We illustrate this feature in Fig.~\ref{fig:6}, where we plot the time evolution of the concurrence for two different arrangements of the coupling of the atoms to an external reservoir. It is seen from Fig.~\ref{fig:6} that the concurrence of the approximate state (\ref{eq:50}) coincides quite well with the concurrence of the actual state of the system. The reason for this similarity is that, in the limit of $\gamma_{12}=0$, the symmetric and antisymmetric states are equally populated by the spontaneous emission from the upper state~$\ket e$. This results in an equal redistribution of the noise between the symmetric and antisymmetric states. In other words, the antisymmetric state is now fully participating in the dynamics of the system, and the actual state of the system of atoms coupled to independent reservoirs can be well approximated by the density matrix \eqref{eq:50}.  

We close this part of the section by pointing out that following the results presented in Figs.~\ref{fig:4}-\ref{fig:6} one would conclude that in the presence of spontaneous emission, the entanglement created by the two-photon coherence may exist only over a short time of the evolution of the system. However, a quick look at the figures reveals that the time range over which the entanglement exists depends on the value of the parameters involved. A question then arises, which of the parameters are crucial for the entanglement to survive over a long time, presumably until steady state? In order to examine this point more closely, we consider the steady state solution for the density matrix elements from which we then infer conditions for a non-zero steady state entanglement.

\subsection{Steady-state entanglement}

We now proceed to use the steady state solutions \eqref{eq:A7} in order to evaluate the concurrence of the stationary state of the system. To evaluate the concurrence, defined in Eq.~\eqref{eq:46}, we must first find the eigenvalues of the matrix $R$. By taking the explicit form of the matrix $R$ and making use of the steady-state solutions for the density matrix elements, Eq.~\eqref{eq:A7}, we readily find the following expressions for the square roots of the required eigenvalues
\begin{align}
&\sqrt{\lambda_{1,2}} =\frac{\tilde{\Omega}^{2}}{4}\frac{\sqrt{\tilde{\Omega}^{4}
+4\gamma^{2}|U_{12}|^{2}}\pm
2\gamma |U_{12}|} {\tilde{\Omega}^{4}+\gamma^{2}\!\left[2\tilde{\Omega}^{2}+\left(\gamma +\gamma_{12}\right)^{2} +\Omega_{12}^2\right]} ,\nonumber\\
&\sqrt{\lambda_{3}}=\sqrt{\lambda_{4}}=\frac{1}{4}\,\frac{\tilde{\Omega}^{4}}
{\tilde{\Omega}^{4}+\gamma^{2}\!\left[2\tilde{\Omega}^{2}\!+\!\left(\gamma\!+\!\gamma_{12}\right)^{2}\!+\!\Omega_{12}^2\right]} ,\label{eq:57}
\end{align}
where $U_{12} =\gamma_{12} +i\Omega_{12}$ describes the strength of the interaction between the atoms.

Although the above formulas look complicated, a straightforward but lengthy calculation leads to a remarkably simple analytical expression for the steady-state concurrence
\begin{align}
{\cal C}(\infty) = {\rm max}\!\left\{0, \frac{\tilde{\Omega}^{2}\left(\gamma |U_{12}|-\frac{1}{2}\tilde{\Omega}^{2}\right)}{\tilde{\Omega}^{4}+\gamma^{2}\!\left[2\tilde{\Omega}^{2}+\left(\gamma +\gamma_{12}\right)^{2}+\Omega_{12}^2\right]}\!\right\} .\label{eq:96}
\end{align}
Equation (\ref{eq:96}) is a general formula for the steady state concurrence valid for an arbitrary Rabi frequency, an arbitrary distance between the atoms, and for a common $(\gamma_{12}\neq 0)$ or separate $(\gamma_{12}= 0)$ reservoirs. It is seen that there is a threshold for the Rabi frequency below which the atoms are entangled in the steady state. The threshold depends only on the relation between the Rabi frequency and the strength of the interaction between the atoms. The interaction is determined by the collective parameters~$\Omega_{12}$ and~$\gamma_{12}$.

In the special case of independent reservoirs, i.e., when $\gamma_{12}=0$, the concurrence reduces to the result recently obtained by Li and Paraoanu~\cite{lp09}
\begin{align}
  \label{eq:97}
  {\cal C}(\infty) = {\rm max}\left\{0, \frac{\tilde{\Omega}^{2}\left(\gamma\Omega_{12} -\frac{1}{2}\tilde{\Omega}^{2}\right)}{(\tilde{\Omega}^{2}+\gamma^{2})^{2}+\gamma^{2}\Omega_{12}^{2}}\right\} ,
\end{align}
which also exhibits a threshold for the steady-state entanglement. The existence of a threshold was also noted by Macovei {\it et al.}~\cite{me10}, who studied numerically the stationary pairwise entanglement in a system composed of $N$ atoms confined to a region much smaller than the resonant wavelength, i.e., in the limit of~$\gamma_{12}=\gamma$ and $\rho_{aa}^{s}=0$.

The steady-state concurrence \eqref{eq:96} depends on the value of the Rabi frequency of the driving field relative to strength of the interatomic interactions. Obviously, the stationary entanglement is zero for independent atoms, i.e., $\gamma_{12}=0$ and $\Omega_{12}=0$. However, as soon as $\gamma_{12}\neq 0$ and/or $\Omega_{12}\neq 0$, the atoms can be entangled in the steady state. The sufficient condition for a steady state entanglement is to maintain the strength of the interatomic interactions $\gamma |U_{12}| > \tilde{\Omega}^{2}/2$. Thus, by changing the value of $\tilde{\Omega}$, we may dynamically switch on or switch off the steady state entanglement. 

Looking at the steady-state solutions for the density matrix elements, Eq.~\eqref{eq:A7}, we see that in the limit of a strong dipole-dipole interaction, $\Omega_{12}\gg \gamma$, the two-photon coherence $\rho_{ge}$ approaches a large non-zero value, $\rho_{ge}(\infty)\approx -i/5$ with all other coherences being vanishingly small. Evidently, the source of the two-photon coherence is in the dipole-dipole interaction that shifts the single excitation states from the resonance with the driving laser field. The strong correlations are reflected in the stationary state of the atoms which despite their interaction with a dissipative reservoir, decay to a strongly correlated state. 
\begin{figure}[h]
  \includegraphics[width=0.95\columnwidth]{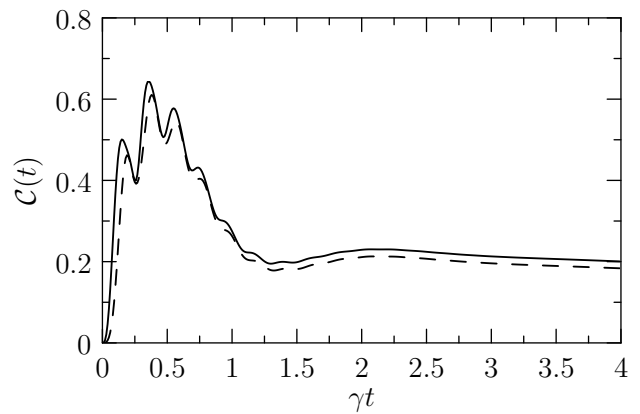}
  \caption{The time evolution of the concurrences ${\cal C}(t)$ (solid line) and ${\cal C}_{x}(t)$
    (dashed line) for $\Omega_{0} =10\gamma$, $\theta =\pi/2$ and $r_{12}=0.06\lambda$, ($\gamma_{12}=0.97\gamma$, $\Omega_{12}=26.22\gamma$).}
  \label{fig:11}
\end{figure}

The existence of the threshold for the steady-state entanglement provides a clear explanation of why for the examples illustrated in Figs.~\ref{fig:4}-\ref{fig:6} the entanglement was confined to short times. Simply, the chosen values of the parameters corresponded to the below threshold situations of the interatomic interaction strength $\gamma |U_{12}| < \tilde{\Omega}^{2}/2$. If instead, we choose the values of the parameters such that $\gamma |U_{12}| > \tilde{\Omega}^{2}/2$, i.e. above the threshold, we then should observe an entanglement for all times $t>0$. This is illustrated in Fig.~\ref{fig:11}, where we plot the concurrences~${\cal C}(t)$ and ${\cal C}_{x}(t)$ for the case of $\gamma |U_{12}| > \tilde{\Omega}^{2}/2$. The concurrence builds up in an oscillatory fashion and then approaches a non-zero steady-state value~${\cal C}(\infty) \approx 0.2$. Clearly, the entanglement is present for all times $t>0$. It is also seen that the concurrence of the $X$ state admits of lower level of entanglement than the actual state of the system, i.e., ${\cal C}_{x}(t)<{\cal C}(t)$ for all times.

\section{Potential experimental scheme}

Finally, we discuss experimental prospects for the realisation of the two-photon Bell states in a system of two driven atoms. A potential experimental system might be based on the experimental setup of Ga\"etan {\it et al.}~\cite{ge10,gm09} and the same set of parameters employed in the observation of the Rydberg blockade effect. 
The only difference would be in the tuning of the excitation laser field to the energy levels of the system. Namely,  instead of satisfying the condition of zeroing the detuning between the ground state $\ket g$ and the single-excitation entangled state $\ket s$, the frequency of the excitation laser should be kept on the two-photon resonance between the ground state~$\ket g$ and one of the upper states shifted by the dipole-dipole interaction potential $\Delta E_{\pm}=\pm C_{3}/r_{12}^{3}$, where~$C_{3}$ is a constant depending on the geometry of the experiment~\cite{ge10,gm09}. In this case, the frequency shift $\Delta\omega_{\pm} =\Delta E_{\pm}/\hbar$ appears as a detuning of the laser field from the one-photon resonance $\ket g \rightarrow \ket s$. In other words, a frequency shift $\Delta \omega_{\pm} >\gamma$ would prevent the excitation of the state $\ket s$. A practical value of the dipole-dipole frequency shift $\Delta\omega_{\pm}\approx 50$ MHz and a typical spontaneous emission rate of highly excited states of $^{87}$Rb atoms of $\gamma/2\pi \sim 6$ MHz satisfy the requirement of $\Delta \omega_{\pm} >\gamma$. 
One could argue that such a scheme differs from that considered in this paper and illustrated in Fig.~\ref{dfig1}. 
However, these two schemes are mathematically completely equivalent to each other. It is seen from Eq.~\eqref{eq:A4} that the dipole-dipole interaction appears in the equations of motions for the coherences $\rho_{gs}$ and $\rho_{se}$ as a detuning of the laser field from the resonance frequency of the transition $\ket g \rightarrow \ket s$. The detuning $\Omega_{12}$ can be achieved either through a shift of the entangled state $\ket s$ from the atomic resonance by $\Omega_{12}$ or through a shifting of the laser field frequency from the atomic resonance by the amount of $\Omega_{12}$.  

In closing, we briefly comment about a possible experimental observation of the signature of entanglement created by the two-photon coherence. We have seen that the entangled state created by the shift of the single excitation states involves states with zero and double excitations. These states are intrinsically connected to correlated states known in the literature as pairwise atomic squeezed states~\cite{bd87,ap90,zf91}. Similar states occur at long times for a two-atom system in a squeezed reservoir. Therefore, a signature of such a state should be seen in squeezing of the emitted fluorescence field. Hence, the presence of the entangled states could be detected simply by observing squeezed fluctuations in one of the quadratures of the emitted fluorescence field. The fluctuations are directly measurable in schemes involving homodyne or heterodyne detection. 
\begin{figure}[h]
  \includegraphics[width=0.95\columnwidth]{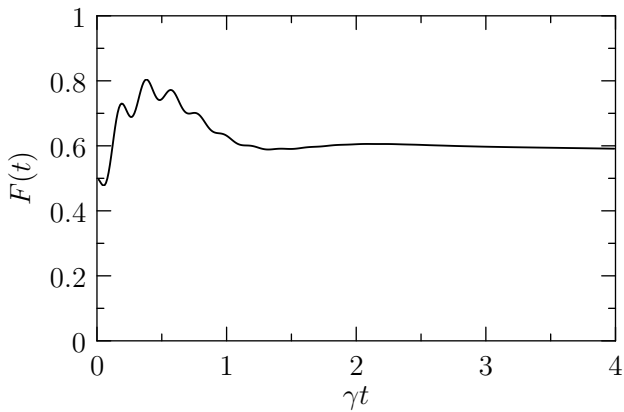}
  \caption{The time variation of the fidelity $F(t)$ for $\Omega=10\gamma$, $\theta =\pi/2$ and $r_{12}=0.06\lambda$, ($\gamma_{12}=0.97\gamma$, $\Omega_{12}=26.22\gamma$).}
  \label{fig:12}
\end{figure}

Alternatively, one could measure fidelity of the entangled state created by the two-photon coherence in the presence of spontaneous emission. It could be done in experiments similar to that of Refs.~\cite{zi10,wg10}, where the fidelity of entangled single excitation states was observed. Fidelity of the actual mixed state of the system, determined by the density matrix (\ref{eq:49}), is given by 
\begin{eqnarray}
F(t) = \bra{\Psi}\rho(t)\ket{\Psi} = \frac{1}{2}\!\left[\rho_{ee}(t)\!+\!\rho_{gg}(t)\right]\!+\!{\rm Im}[\rho_{eg}(t)] ,
\end{eqnarray}
where $\ket{\Psi}$ is the pure two-photon maximally entangled Bell state, Eq.~\eqref{eq:38}, created by the two-photon coherence in the absence of spontaneous emission. 

Figure~\ref{fig:12} shows a plot of the fidelity versus the normalised time $\gamma t$ given by this equation for the same parameters as in Fig.~\ref{fig:11}.  
We see that for all times, the fidelity of the actual state is smaller than one, but the essential point is that the concurrence exceeds the value of~$0.5$ required for quantum entanglement. The fidelity is greatest in the transient regime, where it reaches the maximum value $F(t)\approx 0.8$, and then decreases to its steady-state value $F(t)\approx 0.6$. Therefore, the entanglement created by the two-photon coherence is not washed out by spontaneous emission and should be observable in practice.

\section{Conclusions}\label{sec6}

We have presented a mechanism for a controlled generation of pure or mixed Bell states with correlated atoms that involve double or zero excitations. The mechanism inhibits excitation of singly excited collective states of a two-atom system by shifting the states from the one-photon resonance. In particular, we have shown that the shift of the energy levels can lead the system to evolve into a pure entangled Bell state that involves two-atom states with double or zero excitations. The crucial for the occurrence of the entangled state is the presence of a non-zero two-photon coherence. The degree of entanglement and the purity of the state depend on the relaxation of the atomic excitation. In the absence of the atomic relaxation, the state of the system evolves harmonically between a separable to the maximally entangled Bell state. We have found that the concurrence can be different from zero only in the presence of the dipole-dipole interaction. By going into the limit of a large dipole-dipole interaction, we have shown that the concurrence reduces to that predicted for an $X$-state of the system. Furthermore, we have demonstrated that the concurrence of an $X$-state system is a lower bound for the concurrence of the two-atom system. 
In the presence of the relaxation, the general state of the system is a mixed state that under a strong dipole-dipole interaction reduces to an $X$-state form. We have found that mixed states admit of lower level of entanglement, and the entanglement may occur over a finite range of time. The time range for the entanglement depends on the relation between the dipole-dipole interaction and the Rabi frequency of the laser field. We have calculated the steady state concurrence and have found there is a threshold value for the dipole-dipole interaction relative to the Rabi frequency above which the atoms can be entangled for all times.

\section*{ACKNOWLEDGMENTS}
This work was supported in part by a research grant from the King Abdulaziz City for Science and Technology.

\appendix

\section{}

The matrix elements of the atomic density operator $\rho$ written in the basis of the collective states of the system satisfy two independent sets of differential equations. This is a consequence of
using the collective basis, and of the fact of assuming that both atoms experience the same amplitude and phase of the driving field. The two sets can be written in compact matrix forms as
\begin{align}
  \dot{\vec{X}} = M\vec{X} +\vec{I} ,\quad \dot{\vec{Y}} = Q\vec{Y} ,\label{eq:A1}
\end{align}
where the vectors $\vec{X}$ and $\vec{Y}$ are of the form
\begin{align}
\vec{X} &= (\rho_{ee},\rho_{ss},\rho_{aa},\rho_{ge},\rho_{eg},\rho_{es},\rho_{se},\rho_{gs},\rho_{sg})^{T} ,\nonumber\\
\vec{Y} &= (\rho_{ae},\rho_{ea},\rho_{ga},\rho_{ag},\rho_{sa},\rho_{as})^{T} ,\label{eq:A2}
\end{align}
the vector $\vec{I}$ of the inhomogeneous terms is of the form
\begin{align}
\vec{I} = (0,0,0,0,0,0,0,-i\tilde{\Omega},i\tilde{\Omega})^{T} ,\label{eq:A3}
\end{align}
and $M$ and $Q$ are, respectively, $9\times 9$ and $6\times 6$ matrices of the coefficients of the differential equations. 

The equations of motion for the nine components of the vector $\vec{X}$ are
\begin{align}
\label{eq:A4}
  \dot{\rho}_{ee} =& -4\gamma\rho_{ee}+i\tilde{\Omega}(\rho_{se}-\rho_{es}) ,\nonumber\\
  \dot{\rho}_{ss} =& -2(\gamma+\gamma_{12})(\rho_{ss}-\rho_{ee})\nonumber\\
  &+i\tilde{\Omega}(\rho_{es}-\rho_{se}+\rho_{gs}-\rho_{sg}) ,\nonumber\\
  \dot{\rho}_{aa} =& -2(\gamma-\gamma_{12})(\rho_{aa}-\rho_{ee}) ,\nonumber\\
  \dot{\rho}_{ge} =& -2\gamma\,\rho_{ge} +i\tilde{\Omega}(\rho_{se}-\rho_{gs}) ,\nonumber\\
  \dot{\rho}_{eg} =& -2\gamma\,\rho_{eg}-i\tilde{\Omega}(\rho_{es}-\rho_{sg}) ,\nonumber\\
  \dot{\rho}_{es} =& -(3\gamma+\gamma_{12}-i\Omega_{12})\rho_{es}
  +i\tilde{\Omega}(\rho_{ss}-\rho_{ee}-\rho_{eg}) ,\nonumber\\
  \dot{\rho}_{se} =& -(3\gamma+\gamma_{12}+i\Omega_{12})\rho_{se}
  -i\tilde{\Omega}(\rho_{ss}-\rho_{ee}-\rho_{ge}) ,\nonumber\\
  \dot{\rho}_{gs} =& -i\tilde{\Omega} -(\gamma+\gamma_{12}-i\Omega_{12})\rho_{gs}+2(\gamma+\gamma_{12})\rho_{se}\nonumber\\
  &+i\tilde{\Omega}(2\rho_{ss}+\rho_{aa}+\rho_{ee}-\rho_{ge}) ,\nonumber\\
  \dot{\rho}_{sg} =& \, i\tilde{\Omega} -(\gamma+\gamma_{12}+i\Omega_{12})\rho_{sg}+2(\gamma+\gamma_{12})\rho_{es}\nonumber\\
  &-i\tilde{\Omega}(2\rho_{ss}+\rho_{aa}+\rho_{ee}-\rho_{eg}) ,
\end{align}
and the equations of motion for the six components of the vector $\vec{Y}$ are
\begin{align}
\dot{\rho}_{ae}&=-(3\gamma-\gamma_{12}-i\Omega_{12})\rho_{ae}-i\tilde{\Omega}\rho_{as} ,\nonumber\\
\dot{\rho}_{ea}&=-(3\gamma-\gamma_{12}+i\Omega_{12})\rho_{ea}+i\tilde{\Omega}\rho_{sa} ,\nonumber\\
\dot{\rho}_{ga}&=-(\gamma-\gamma_{12}+i\Omega_{12})\rho_{ga}-2(\gamma-\gamma_{12})\rho_{ae}+i\tilde{\Omega}\rho_{sa} ,\nonumber\\
\dot{\rho}_{ag}&=-(\gamma-\gamma_{12}-i\Omega_{12})\rho_{ag}-2(\gamma-\gamma_{12})\rho_{ea}-i\tilde{\Omega}\rho_{as} ,\nonumber\\
\dot{\rho}_{sa} &= -2(\gamma+i\Omega_{12})\rho_{sa} +i\tilde{\Omega}(\rho_{ea}+\rho_{ga}) ,\nonumber\\
\dot{\rho}_{as} &= -2(\gamma-i\Omega_{12})\rho_{as} -i\tilde{\Omega}(\rho_{ae}+\rho_{ag}) ,\label{eq:A5}
\end{align}
where $\tilde{\Omega} =\Omega_{0}/\sqrt{2}$.

The matrices $M$ and $Q$ are non-singular, so we can readily solve the equations (\ref{eq:A1}) by matrix inversion. A formal integration gives
\begin{align}
  \vec{X}(t) &= \vec{X}(0)\exp(Mt) + \left[\exp(Mt)-1\right]\vec{I} ,\nonumber\\
  \vec{Y}(t) &= \vec{Y}(0)\exp(Qt) ,\label{eq:A6}
\end{align}
where $\vec{X}(0)\equiv \vec{X}(t=0)$ and $\vec{Y}(0)\equiv \vec{Y}(t=0)$ are vectors of initial values of the matrix elements.

It is easy to see from Eqs.~(\ref{eq:A4}) and (\ref{eq:A5}) that only the components of the vector~$\vec{X}$ can have nonzero steady-state solutions. All the components of the vector $\vec{Y}$ are zero in the steady-state. 

After straightforward but quite tedious calculations, we find that the steady-state solution for the components of the vector~$\vec{X}$ are 
\begin{align}
\rho_{ee}^{s} =\rho_{aa}^{s} &= \frac{1}{4}\frac{\tilde{\Omega}^{4}}{D} ,\nonumber\\
\rho_{ss}^{s} &= \frac{1}{4}\frac{\left(\tilde{\Omega}^{2} +4\gamma^{2}\right)\tilde{\Omega}^{2}}{D} ,\nonumber\\
\rho_{ge}^{s} = \left(\rho_{eg}^{s}\right)^{\ast} &= -\frac{1}{2}\frac{\gamma\left(\gamma +U_{12}\right)\tilde{\Omega}^{2}}{D} ,\nonumber\\
\rho_{es}^{s} = \left(\rho_{se}^{s}\right)^{\ast} &= \frac{i}{2}\frac{\gamma\tilde{\Omega}^{3}}{D} ,\nonumber\\
  \rho_{gs}^{s} =\left(\rho_{sg}^{s}\right)^{\ast} &= -\frac{i}{2}\frac{\gamma\tilde{\Omega}\left[\tilde{\Omega}^{2}+2\gamma\left(\gamma + U_{12}\right)\right]}{D} ,\label{eq:A7}
\end{align}
where the superscript $"s"$ stands for the steady-state value, 
\begin{align}
  D = \tilde{\Omega}^{4}+\gamma^{2}\left[2\tilde{\Omega}^{2}+\left(\gamma +\gamma_{12}\right)^{2} +\Omega_{12}^{2}\right] ,\label{eq:A8}
\end{align}
and $U_{12} = \gamma_{12} + i\Omega_{12}$.

\end{document}